\documentstyle[psfig,astrobib_ale]{mn}
 
\def\be{\begin{equation}}
\def\ee{\end{equation}}
\def\ni{\noindent}
\def\bfig{\begin{figure}}
\def\efig{\end{figure}}
\def\bfigw{\begin{figure*}}
\def\efigw{\end{figure*}}
\def\etal{et al.~}
\def\etalc{et al.}
\def\eg{{e.g.}}
\def\ie{{i.e.}}
\def\vs{{\em vs}}
\def\mc{\multicolumn}
\def\spose#1{\hbox to 0pt{#1\hss}}
\def\approxlt{\mathrel{\spose{\lower 3pt\hbox{$\sim$}}
	\raise 2.0pt\hbox{$<$}}}
\def\approxgt{\mathrel{\spose{\lower 3pt\hbox{$\sim$}}
	\raise 2.0pt\hbox{$>$}}}
\def\aap{A\&A}

\def\apj{ApJ}
\def\apjl{ApJL}
\def\apjs{ApJS}
\def\araa{ARA\&A}
\def\mnras{MNRAS}

\def\ROSAT{{\em ROSAT }}
\def\PSPC{{\em PSPC }}
\def\pspc{{\em PSPC }}
\def\HRI{{\em HRI }}
\def\Chandra{{\em Chandra }}
\def\ISO{{\em ISO }}
\def\ISOc{{\em ISO}}

\def\ROSATc{{\em ROSAT}}
\def\IRAS{{\em IRAS }}
\def\cm{{\rm\thinspace cm}}

\def\erg{{\rm\thinspace erg}}

\def\mJy{{\rm\thinspace mJy}}
\def\uJy{{\rm\thinspace \mu Jy}}

\def\MHz{{\rm\thinspace MHz}}
\def\GHz{{\rm\thinspace GHz}}

\def\keV{{\rm\thinspace keV}}

\def\km{{\rm\thinspace km}}

\def\Mpc{{\rm\thinspace Mpc}}

\def\s{{\rm\thinspace s}}
\def\ks{{\rm\thinspace ks}}

\def\Myr{{\rm\thinspace Myr}}
\def\mag{{\rm\thinspace mag}}

\def\col{\hbox{$\cm^{-2}$}}
\def\fluxerg{\hbox{$\erg\cm^{-2}\s^{-1}\,$}}

\def\ergps{\mbox{$\erg\s^{-1}$}}

\def\kmps{\hbox{$\km\s^{-1}\,$}}
\def\kmpspMpc{\hbox{$\km\s^{-1}\Mpc^{-1}\,$}}
\def\um{\hbox{$\mu {\rm m}$}}
\def\Ang{{\rm\thinspace \AA}}
\def\bband{$B$--\,band~}
\def\vband{$V$--\,band~}
\def\rband{$R$--\,band~}
\def\kband{$K$--\,band~}
\def\bbandc{$B$--\,band}

\def\rbandc{$R$--\,band}
\def\kbandc{$K$--\,band}
\def\mch{M$\rm^{c}$Hardy}
\def\alpharsm{\alpha^{350}_{1.4}}
\def\Ha{{\rm H}\alpha}
\def\Hb{{\rm H}\beta}

\begin{document}
 
\title[Starburst activity in a \ROSAT NELG]
{Starburst activity in a \ROSAT Narrow Emission-Line Galaxy} 
\author[K. F. Gunn \etal]
{Katherine F. Gunn$^1$\thanks{email: {\tt kfg@astro.soton.ac.uk}},
I.M.\mch$^1$, O.Almaini$^2$, T.Shanks$^3$, T.J.Sumner$^4$, 
\newauthor T.W.B.Muxlow$^5$, A.Efstathiou$^4$, 
L.R.Jones$^6$, S.M.Croom$^7$, J.C.Manners$^2$,
\newauthor A.M.Newsam$^8$, K.O.Mason$^9$, 
S.B.G.Serjeant$^4$, M.Rowan-Robinson$^4$.
\\
$^1$Department of Physics \& Astronomy, University of Southampton,
Highfield, Southampton, SO17 1BJ.  \\
$^2$Institute for Astronomy, The University of Edinburgh, Royal
Observatory, Blackford Hill, Edinburgh, EH9 3HJ.\\
$^3$Department of Physics, University of Durham, South Road, Durham,
DH1 3LE. \\
$^4$Imperial College of Science, Technology and Medicine, Blackett
Laboratory, Prince Consort Road, London, SW7 2BZ. \\
$^5$University of Manchester, Nuffield Radio Astronomy Laboratories,
Jodrell Bank, Macclesfield, Cheshire, SK11 9DL. \\
$^6$School of Physics \& Astronomy, University of Birmingham
Edgbaston, Birmingham, B15 2TT. \\
$^7$Anglo-Australian Observatory, PO Box 296, Epping, NSW 2121,
Australia. \\
$^8$Astrophysics Research Institute, Liverpool John Moores University,
Twelve Quays House, Egerton Wharf, Birkenhead, CH41 1LD. \\
$^9$Mullard Space Science Laboratory, University College London,
Holmbury St Mary, Dorking, RH5 6NT.}
 
\date{DRAFT -- DO NOT DISTRIBUTE}
\pagerange{\pageref{firstpage}--\pageref{lastpage}}
\pubyear{2000}
 
\label{firstpage}

\maketitle
 
\begin{abstract}
We present multiwaveband photometric and optical spectropolarimetric
observations of the $R=15.9$ narrow emission line galaxy R117\_A which
lies on the edge of the error circle of the \ROSAT X-ray source R117
(from \citeNP{McHardy98}).  The overall spectral energy distribution
of the galaxy is well modelled by a combination of a normal spiral
galaxy and a moderate-strength burst of star formation.  The far
infra-red and radio emission is extended along the major axis of the
galaxy, indicating an extended starburst.

On positional grounds, the galaxy is a good candidate for the
identification of R117 and the observed X-ray flux is very close to
what would be expected from a starburst of the observed far infra-red
and radio fluxes.  Although an obscured high redshift QSO cannot
be entirely ruled out as contributing some fraction of the X-ray flux,
we find no candidates to $K=20.8$ within the X-ray errorbox and so
conclude that R117\_A is responsible for a large fraction, if not all,
of the X-ray emission from R117.

Searches for indicators of an obscured AGN in R117\_A have so far
proven negative; deep spectropolarimetric observations show no signs
of broad lines to a limit of one per cent and, for the observed far
infra-red and radio emission, we would expect a ten times greater X-ray
flux if the overall emission were powered by an AGN.  We therefore
conclude that the X-ray emission from R117 is dominated by starburst
emission from the galaxy R117\_A.

\end{abstract}

 
\section{Introduction}

Despite extensive investigation, there remains much uncertainty
regarding the origin of the X-ray emission from narrow emission line
galaxies (NELGs) found in deep X-ray surveys (\citeNP{CCRSS1};
\citeNP{McHardy98}; \citeNP{Schmidt98}; \nocite{Roche95}
\citeANP{Roche96} 1995, 1996).  There are several possible emission
mechanisms, where the two most plausible are starburst activity or the
presence of an active galactic nucleus (AGN).  A burst of intense
star-forming activity creates large numbers of hot OB stars, which are
strong X-ray emitters, and such galaxies have been proposed as a
significant contributor to the X-ray Background (XRB) Radiation
(\eg~\citeNP{GP90}).  It is also generally agreed that obscured AGN
can account for the majority of the energy in the XRB
(\nocite{M94}Madau, Ghisellini \& Fabian 1994; \citeNP{C95}).  AGN
have intrinsically soft spectra, but photo-electric absorption by gas
surrounding the AGN has the effect of preferentially removing lower
energy photons, thereby hardening the observed spectra to agree with
the spectrum of the XRB.

Results from deep \ROSAT X-ray surveys show that the number of X-ray
sources at faint fluxes continues to rise faster than the contribution
from QSOs and clusters (\citeNP{Ioannis96};
\citeNP{McHardy98}).  Through accurate positions from the \ROSAT {\em
HRI} \cite{Has98}, many of these faint X-ray sources appear to be
unambiguously identified with galaxies showing narrow emission lines
in their optical spectra.  It is difficult to determine the dominant
power source from the usual line ratio diagnostics \cite{Osterbrock89}
since many of these objects lie on the border between AGN and
starburst, and it is also likely that the two phenomena may coexist in
some objects (\citeNP{Iwasawa97}; \citeNP{McHardy98}).

Here we present multiwavelength imaging and photometry of R117, a
\ROSAT source from the UK Deep Survey of \citeANP{McHardy98} (1998), 
identified with a NELG at $z=0.061$.  These observations include
spectropolarimetry, to search for broad emission lines in polarized
light as a sign of AGN activity.  In addition, submillimetre (sub-mm)
photometry aims to detect thermal emission from dust associated with
the source of the X-ray flux, whether due to starburst activity within
the galaxy or the presence of an obscured AGN.  The available data
point towards a pure starburst origin to the soft X-ray emission in
R117, which implies that star-forming galaxies may make a significant
contribution to the soft X-ray background radiation.

\section{Observations}
\label{sec:obs}

\begin{table*}
\caption
{Measured flux densities and upper limits for the \ROSAT source R117
and its optical counterpart, galaxy R117\_A, from X-ray to radio
wavelengths. }
\begin{center}
\begin{tabular}{llccll}\hline
Band     & \mc{1}{c}{Wavelength}& Frequency          & Flux density     & Telescope/            & Notes \\
         & \mc{1}{c}{($\um$)}   & (Hz)               & ($\mJy$)         &  ~~~~Instrument       &          \\
\hline
$1\keV$  & $1.24\times10^{-3}$  & $2.4\times10^{17}$ & $8.3\pm2.1\times10^{-7}$ & \ROSATc/{\em PSPC}    & $S$(0.5-2\,keV)=(2.81$\pm$0.70)$\times10^{-15}\fluxerg$ \\
$B$      & $0.44$               & $6.8\times10^{14}$ & $0.595^{+1.00}_{-1.20}$  & INT/WFC               & $B_{\rm tot}=17.14\pm0.2$ \\
$V$      & $0.55$               & $5.5\times10^{14}$ & $1.01^{+0.20}_{-0.17}$   & CFHT                  & $V_{\rm tot}=16.4\pm0.2$ \\
$R$      & $0.70$               & $4.3\times10^{14}$ & $1.20^{+0.24}_{-0.20}$   & UH88                  & $R_{\rm tot}=15.94\pm0.2$$^*$ \\
$K$      & $2.20$               & $1.4\times10^{14}$ & $2.97^{+0.60}_{-0.50}$   & UKIRT/UFTI            & $K_{\rm tot}=13.30\pm0.2$ \\
$12\um$  & $12.0$               & $2.5\times10^{13}$ & $<61$            & \IRAS                 & estimated from FSC sources nearby \\
$15\um$  & $15.0$               & $2.0\times10^{13}$ & $7.4\pm 2.7$
& \ISOc/{\em CAM}       & $14.8\pm1.0$adu ($\pm30\%$ error in mJy/adu conversion) \\
$25\um$  & $25.0$               & $1.2\times10^{13}$ & $<60$            & \IRAS                 & estimated from FSC sources nearby \\
$60\um$  & $60.0$               & $5.0\times10^{12}$ & $200 \pm 35$     & \IRAS                 & SCANPI measurement$^\dagger$ \\
$90\um$  & $90.0$               & $3.3\times10^{12}$ & $210 \pm 70$     & \ISOc/{\em PHOT}      & includes calibration errors \\
$100\um$ & $100.0$              & $3.0\times10^{12}$ & $270 \pm 64$     & \IRAS                 & SCANPI measurement$^\dagger$ \\
$450\um$ & $450.0$              & $6.7\times10^{11}$ & $<24.9$          & JCMT/SCUBA            & $3\sigma$ upper limit \\
$850\um$ & $850.0$              & $3.5\times10^{11}$ & $4.01\pm 0.99$   & JCMT/SCUBA            & \\
$6\cm$& $6.0\times10^4$     & $4.9\times10^{9}$  & $0.45 \pm 0.05$      & VLA                   & \\
$20\cm$& $2.0\times10^5$     & $1.4\times10^{9}$  & $1.4 \pm 0.05$      & VLA                   & \\
\hline
\end{tabular}
\end{center}
\flushleft
{\small $^*$The magnitude given here is the total magnitude of the
galaxy. The magnitudes given in \citeANP{McHardy98} are from 6 arcsec
apertures which slightly underestimate the flux from large extended
galaxies such as R117\_A. \\
$^\dagger$\IRAS fluxes obtained using the SCANPI facility at IPAC, Caltech.}
\label{tab:data}
\end{table*}

In this Section, we present the multiwavelength data obtained for the
\ROSAT source, R117, and the proposed optical counterpart.  Each of
the observations of R117 is described in turn, from the X-ray through
to radio wavelengths, and the measurements are summarized in Table
\ref{tab:data}.  We take ${\rm H}_0 = 75 \kmpspMpc$ and $q_0 = 0.5$
throughout.

\subsection{X-ray Data}

\ROSAT X-ray source 117, henceforth known as R117, was detected in the
UK \ROSAT Deep Survey \cite{McHardy98}.  The UK \ROSAT Deep Survey
consists of \pspc observations totalling $115\ks$ at the position
$13^{\rm h} 34^{\rm m} 37\fs0 +37^\circ 54' 44''$ (J2000), which lies
in a region of low Galactic absorption, $N_H \sim 6.5 \times 10^{19}
\col$.  A complete sample of 70 X-ray sources is defined above a flux
limit of $S(0.5-2\keV) = 2 \times 10^{-15} \fluxerg$.  R117 is one of
the fainter sources in the sample, with a flux of $S(0.5-2\keV) =
(2.81 \pm 0.70) \times 10^{-15} \fluxerg$, so the measurement of its
X-ray spectrum is not possible. R117 is not detected in the soft
($0.1-0.5\keV$) band, putting a limit on its hardness ratio, \ie, the
ratio of counts in the $0.5-2\keV$ band to those in the $0.1-0.5\keV$
band, of $HR > 0.41$.  This value of the hardness ratio would
correspond to an unabsorbed power-law of energy spectral index $\alpha
\la 1.2$ (c.f. table 6 in \citeNP{McHardy98}), where $S(\nu) \propto
\nu^{-\alpha}$.

\subsection{Selection of the Optical Counterpart}
\label{sec:select}

\bfigw
\begin{minipage}{3in}
\psfig{figure=r117_b_con.ps,width=2.9in,angle=270} 
\end{minipage} 
\hspace*{0.1in}
\begin{minipage}{3in}
\psfig{figure=r117_r_con.ps,width=2.9in,angle=270}
\end{minipage} 
\begin{minipage}{3in}
\vspace*{0.2in}
\psfig{figure=r117_k_ufti_smooth_con.ps,width=2.9in,angle=270}
\end{minipage} 
\hspace*{0.1in}
\begin{minipage}{3in}
\vspace*{0.2in}
\psfig{figure=r117_radio.ps,width=2.9in,angle=270} 
\end{minipage} 
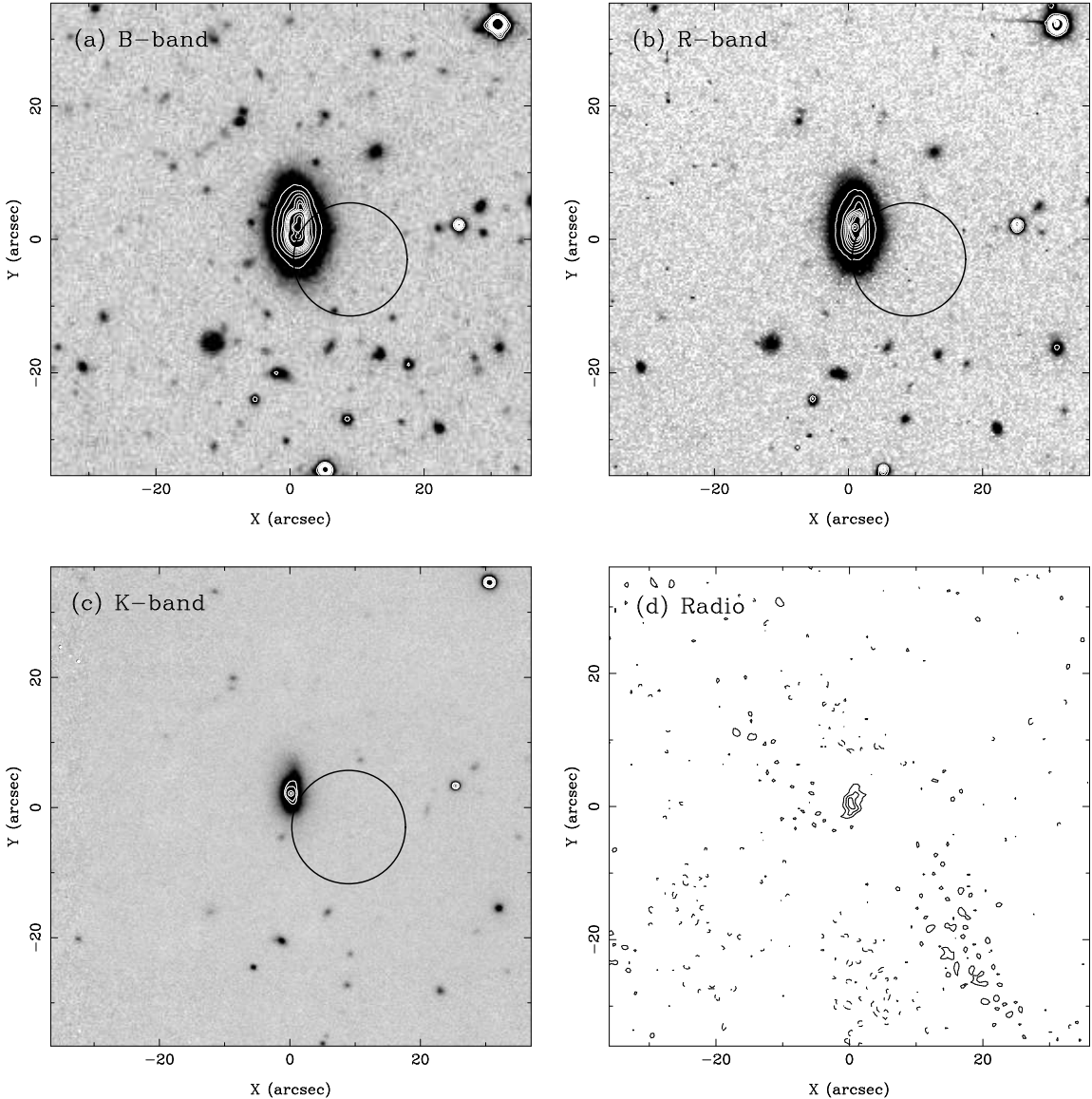
\caption{$B$--, $R$-- and \kband imaging of R117, to the same scale,
showing the position of the {\em ROSAT} error circle ($\sim 10''$
radius) in each case.  (a) \bbandc: INT Wide Field Camera, (b)
\rbandc: CFHT Luppino camera, (c) \kbandc: UFTI on UKIRT, and (d)
$1.4\GHz$ radio map, taken with the VLA A-array. The contour levels
are at $-34, 34, 69$ and $103 \uJy$/beam.}
\label{fig:phot}
\efigw

The potential counterparts to the X-ray sources in the UK \ROSAT Deep
Survey are found from deep optical CCD images.  In
Fig.~\ref{fig:phot}(a-c), we present the available digital imaging
data for R117, in the optical and near infra-red.

Our simulations show that, for an X-ray source with the flux of R117,
the 95 per cent error-radius is $\sim 10$ arcsec (fig.~4 of
\citeNP{McHardy98}).  We therefore overlay the 10 arcsec X-ray circle on
each of these three images, from which it can clearly be seen that the
most obvious optical counterpart to the X-ray source is a bright
galaxy ($13^{\rm h} 34^{\rm m} 13\fs55 +37^\circ 45' 39\farcs0$ J2000)
which lies only 11.9 arcsec from the \PSPC X-ray centroid ($13^{\rm h}
34^{\rm m} 12\fs61 +37^\circ 45' 35\farcs0$ J2000).  This galaxy,
referred to here as R117\_A, is therefore positionally a reasonable
candidate.  We should however consider the possibility of a galaxy of
that magnitude ($R=15.9$) occurring at that position by chance.  The
surface density of galaxies of $R\leq16$ is 0.017 per square arcmin
\cite{Jones91}. There are 30 errorboxes in \citeANP{McHardy98} which
do not have an unambiguous identification with a QSO, a cluster of
galaxies or a star.  For an assumed average errorbox radius of 10
arcsec, we therefore expect 0.04 random occurences with galaxies of
$R\leq16$.  If we take a rather pessimistic average errorbox radius of
15 arcsec, then we double the number of expected chance coincidences,
giving a 0.3 per cent probability of such a galaxy being in the R117
errorbox by chance.  Of course, if we consider only starburst
galaxies, then this probability is reduced still further.

In view of the claim by \citeANP{Lehmann2000} (2000) that all of the
low luminosity NELGs in the UK Deep Survey are chance coincidences, we
must consider carefully the likelihood that this galaxy is the correct
identification.  The possibility of an alternative optical counterpart
for the X-ray emission is discussed fully in Section \ref{sec:altopt}.
However, for the remainder of Section \ref{sec:obs} and the whole of
Section \ref{sec:models} we proceed on the assumption that the bright
galaxy, R117\_A, is the correct counterpart.

\subsection{Optical and near infra-red photometry}
\label{sec:brkphot}

From the images in Fig.~\ref{fig:phot}, we obtain total galaxy
magnitudes for R117\_A of $B = 17.14 \pm 0.2$, $R = 15.94 \pm 0.2$ and
$K = 13.30 \pm 0.2$.  Additional \vband photometry, $V=16.4$, gives
$V-R \sim 0.5$, which is more typical of a spiral than an elliptical
galaxy.  A distance modulus of $m-M=36.97$ gives an absolute magnitude
of $M_V = -20.6$.  The surface brightness profile of the \rband image
is well fit by an exponential disc, as expected for a spiral galaxy,
and has a scale length of $\sim 3$ arcsec or $\sim 3 \, {\rm kpc}$ at
the redshift of the galaxy.  The central surface brightness of the
galaxy is $\mu_R \simeq 19.2 \mag \, {\rm arcsec}^{-2}$.  Careful
visual examination of the \bband image indicates the probable presence
of faint spiral structure, although it is very hard to reproduce such
displays on hard copy.  However, in the contours of the \bband image,
there is an indication of some linear structure, possibly indicating a
central bar.

\subsection{Optical Spectroscopy}

An optical spectrum of R117\_A was taken using the ISIS dual-beam
spectrograph on the WHT, and is presented in fig.\,13 of
\citeANP{McHardy98} (1998).  The spectrum shows narrow Balmer
emission lines, with FWHM $< 1000 \kmps$, plus the forbidden lines,
[O\,{\sc ii}], [N\,{\sc ii}] and [S\,{\sc ii}].  From this, a NELG
identification for R117 is obtained, at a redshift of $z=0.061$.

In order to determine the X-ray emission mechanism for R117,
\citeANP{McHardy98} plotted the standard diagnostic line ratios
of [O\,{\sc iii}]$\lambda 5007$/$\Hb$ against [N\,{\sc ii}]/$\Ha$, and
[O\,{\sc iii}]$\lambda 5007$/$\Hb$ against [S\,{\sc ii}]/$\Ha$
\cite{VO87}, which can be used as a discriminant between starburst 
galaxies and AGN.  Although the [O\,{\sc iii}]$\lambda4959$ line in
R117\_A can just be detected, the dichroic at $\sim 5200 \Ang$ has
attenuated the [O\,{\sc iii}]$\lambda5007$ emission, assuming a standard
ratio of 3:1 for $\lambda 5007$:$\lambda 4959$.  R117\_A lies close to
the boundary between the two emission mechanisms, in the starburst
domain, and the uncertainty on the strength of the [O\,{\sc iii}] line
would have the effect of moving the galaxy more firmly into this
region.

\subsection{Optical Spectropolarimetry}

\bfig
\psfig{figure=specpol.eps,angle=270,width=3.3in}
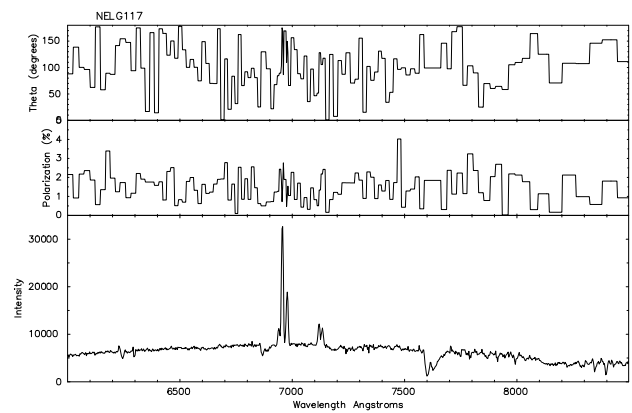
\caption{Polarised spectrum of R117, binned to give a good {\em S/N}
per pixel.  There is a clear detection of continuum polarisation at
about the 1-2 per cent level, but no polarised broad lines.
Spectropolarimetry also yields a very high signal-to-noise total flux
spectrum (lower panel), showing the $\Ha$/N\,{\sc ii}/S\,{\sc ii} region
in detail, with no evidence for any broad wings to the emission lines.
The spectrum is unfluxed, hence the presence of atmospheric absorption
features.  }
\label{fig:specpol}
\efig

Spectropolarimetric observations of a sample of NELGs were taken on
the nights of 1999 June 5-10, using the blue arm of the ISIS
spectrograph on the WHT (Almaini \etal in preparation).  In
Fig.~\ref{fig:specpol}, we present the spectropolarimetric
observations of R117\_A.  The spectrum was calibrated using
observations of both polarized and unpolarized standard stars.
Although a very high $S/N$ spectrum was obtained, and polarization of
the continuum was detected, there was no evidence of polarized broad
emission lines, to a limit of less than one per cent.  This argues
against an AGN origin for the source of the X-ray emission, but cannot
rule it out completely, as any scattering region may also be obscured
\cite{Young96}.

From \nocite{Serkowski75} Serkowski, Mathewson \& Ford (1975), the
maximum contribution to the detected continuum polarization from dust
in our own Galaxy is given by $P_{\rm max} = 9 E(B-V)$.  The line of
sight extinction for R117\_A is $E(B-V) = 0.011$, which limits the
Galactic contribution to $\sim 0.1$ per cent.  This therefore implies
that the detected polarization is intrinsic to R117\_A.  `Normal'
galaxies tend to have continuum polarization at a level of $\la 0.5$
per cent, whereas aligned dust grains in starburst galaxies can cause
a higher level, of the order of the 1-2 per cent continuum
polarization observed in R117\_A (see \eg ~\citeNP{Brindle91}).  The
position angle of the magnetic field vector is at $\sim 10^\circ$,
which is roughly aligned with the major axis of the galaxy, consistent
with the presence of a dust lane or bar.

\subsection{Far Infra-red Data}

The UK Deep \ROSAT field was observed by \ISO as part of the European
Large Area \ISO Survey (ELAIS; \citeNP{Oliver2000}).  Approximately
two-thirds of the \ROSAT field-of-view was covered by
\ISOc, within which 8 \ISO sources were detected at $15\um$.  Of these
8 \ISO sources, 3 are coincident with \ROSAT X-ray sources, and are
identified as 2 stars and 1 galaxy, R117\_A.

R117\_A is the {\it optically} brightest X-ray NELG in the
\ISOc-\ROSAT overlap region, and was detected as an \ISO source at
$15\um$ with a count rate of $14.8\pm1.0 \, {\rm adu}$ using 20 arcsec
radius aperture photometry.  This gives a flux of $S(15\um) = 7.4 \pm
2.7 \mJy$, where the error is mainly due to uncertainty in the
conversion between adu and $\mJy$ (see \citeNP{Serjeant2000}).  The
\ISO $15\um$ image appears extended in the N-S direction, consistent
with the structure seen in the radio data (see Section
\ref{sec:radio}).  At $90\um$, R117\_A is detected and flux calibrated
with the pipeline described in \citeANP{Efstathiou2000} (2000a), with
$S(90\um) = 210 \pm 70 \mJy$.

\subsection{Sub-millimetre Data}

R117 was observed at sub-mm wavelengths as part of a program to
investigate the sub-mm properties of X-ray selected galaxies and QSOs
(Gunn \etal in preparation).  On the night of 1999 May 28, the
Sub-millimetre Common User Bolometer Array (SCUBA; \citeNP{Holland99})
on the JCMT was used in the standard photometry mode to make
simultaneous $850$ and $450\um$ observations.  The secondary mirror
was chopped using a 9 point jiggle pattern, nodding between the signal
and reference beams every $18 \s$.  The data were corrected for the
atmospheric opacity, which was in the range $0.17 < \tau_{850} < 0.22$
during the observation of R117, and calibrated against Mars.  R117 was
detected at $850\um$ with a flux of $S(850\um) = 4.01 \pm 0.99 \mJy$,
whereas at $450\um$ we obtain a $3\sigma$ upper limit of $S(450\um)
\approxlt 24.9 \mJy$.

\subsection{Radio Data}
\label{sec:radio}

Fig.\ref{fig:phot}(d) shows the VLA A-array map of R117 at $1.4\GHz$
($20\cm$).  The A-array data consists of 25 hours of observations from
1999 August 2, 6 and 7, and were made in 4 IF multi-channel continuum
mode.   The map shown in Fig.\ref{fig:phot}(d) is the result of an
initial reduction and residual large scale stripes are still visible
going from north-east to south-west across the map.  However,
well-resolved emission is still easily visible, coincident with the
galaxy.  Interestingly, the spiral structure, which can be discerned
with some difficulty on the optical image, shows up clearly and in
the same location on the radio image.  The total flux density of the
galaxy is $S(1.4\GHz) = 1.4 \pm 0.05 \mJy$.

Observations at $6\cm$ made in full $50 \MHz$ continuum mode with the
VLA on 1991 April 26, reveal a flux of $S(4.9\GHz) = 0.45 \pm 0.05
\mJy$.  These observations indicate a steep ($\alpha \sim 0.95$)
spectral index, implying a starburst origin for the radio emission.

\section{Modelling of the SED}
\label{sec:models}

\bfigw
\centerline{
\psfig{figure=aenufnu.ps,width=4.2in,angle=270}}
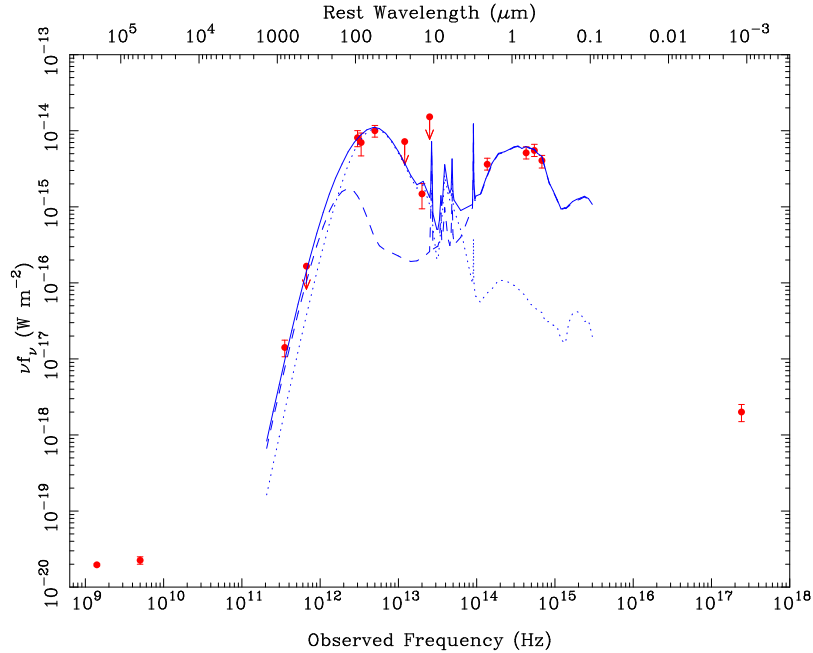
\caption{The spectral energy distribution of R117, showing
all available data and upper limits, from radio to X-ray energies.  We
compare the data to a model fit of an exponentially decaying $17\Myr$
burst of star-formation, with an e-folding time of $20\Myr$ (dotted
line).  The optical light is most likely to be dominated by an older
stellar population, with associated dust, and this cirrus component
(dashed line) contributes about half of the sub-mm flux.  This model
gives a bolometric luminosity of $L_{\rm bol} \sim 5\times10^{10}
L_\odot$.  The data are summarised in Table \ref{tab:data}.}
\label{fig:sed}
\efigw

In Fig.~\ref{fig:sed}, we combine the data described in Section
\ref{sec:obs} to create the spectral energy distribution of R117,
assuming that the bright optical galaxy, R117\_A, is the true
counterpart to the X-ray source.  By plotting $\nu f_\nu$ as a
function of frequency, we see that the bulk of the emission emerges in
two peaks, one in the far infra-red and the other in the optical, with
the far infra-red peak dominating slightly.  The data for R117 (filled
circles) can be fit by the sum (solid line) of two components.  The
bulk of the FIR emission comes from an exponentially decaying
$16.6\Myr$ old burst of star-formation, with an e-folding time of
$20\Myr$ (dotted line).  The optical light is most likely to be
dominated by an older stellar population, with associated dust, and
this cirrus component (dashed line) contributes about half of the
sub-mm flux.  For full details of the models used, see Efstathiou,
Rowan-Robinson \& Siebenmorgen (2000b)\nocite{Efstathiou99}.  This
model gives a bolometric luminosity of $L_{\rm bol} \sim
5\times10^{10} L_\odot$.

\section{Discussion}

So far we have assumed that the bright galaxy, R117\_A, is the true
counterpart of the X-ray source R117.  In this Section, we now
investigate whether the multiwavelength properties of R117 are
consistent with a NELG origin for the X-ray emission, either due to
starburst or AGN activity, or whether we can rule out the optical
counterpart in this way.  We then go on to discuss in detail the
possibilities for alternative sources for the X-ray emission, using
the optical and near infra-red imaging data.

\subsection{X-ray/FIR properties}

\bfig
\centerline{
\psfig{figure=djf92.ps,width=3.0in,angle=270}}
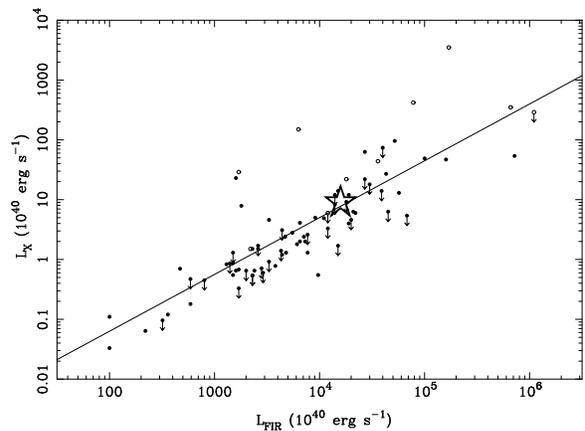
\caption{The far infra-red and X-ray luminosity of R117 (open star)
compared with the samples of David, Jones \& Forman (1992): normal and
starburst galaxies (filled circles), and Seyfert galaxies (open
circles).  Upper limits to the X-ray luminosities are denoted by
arrows.  This shows that R117 lies very close to the best-fit
relationship between $L_X$ and $L_{\rm FIR}$, taken from David \etalc,
for the normal/starburst galaxy sample.}
\label{fig:djf92}
\efig

The combined X-ray and FIR properties of both active and normal
galaxies have been studied by \nocite{Green92}Green, Anderson \& Ward
(1992), who chose a large sample of galaxies from the literature for
which both \IRAS $60\um$ fluxes and {\em Einstein} $0.5-4.5 \keV$
fluxes had been measured.  In this sample, they observe a distinct
bi-modal distribution in the ratio of X-ray to FIR luminosities, with
broad and narrow line galaxies appearing to be distinct populations.
Galaxies with a ratio $L_X/L_{\rm FIR} \approxgt 0.01$ are found to
almost always have broad emission lines.

To find this ratio for R117, we estimate the $0.5-4.5 \keV$ luminosity
from the $0.5-2\keV$ \ROSAT luminosity, by assuming a power-law slope
of $\alpha = 0.5$, typical of NELGs (\citeNP{Romero96};
\citeNP{Almaini96}), which gives $L_X(0.5-4.5\keV) = 9 \times 10^{40}
\ergps$ (c.f $L_X(0.5-4.5\keV) = 7 \times 10^{40} \ergps$ for
$\alpha=1$).  To calculate the FIR luminosity, i.e., the luminosity in
the \IRAS $60\um$ (restframe) band, we use the relationship from Green
\etal (1992), which uses the flux density at $60\um$, $f_{60}$, and
the spectral slope between $25$ and $100\um$, $\alpha(25,100)$.  For
R117, we take the \IRAS $60\um$ flux density, $f_{60} \sim 200
\mJy$, and we use the $15$ and $90\um$ fluxes to estimate 
$\alpha(25,100)$, as being the most reliable measurements.  This gives
a value of $\alpha(25,100) = -2.3$, compared with that measured for
emission line galaxies of $-1.8$, spirals/irregulars of $-1.5$, and
QSOs of $-0.9$.  In this way, we find for R117, $L_{\rm FIR} = 1.8
\times 10^{44} \ergps$.

The X-ray to FIR luminosity ratio for R117, $L_X/L_{\rm FIR} \sim 5.0
\times 10^{-4}$, places R117 firmly in the domain of narrow emission
line galaxies, as expected from the optical data.  Green \etal include
all types of narrow-line galaxies in this class, including Seyfert 2s,
and therefore this method cannot be used as a discriminant between an
AGN and starburst origin for the X-ray and FIR emission.  However,
such a correlation between the $L_X$ and $L_{\rm FIR}$ is to be
expected, given a common stellar origin for both components.

David, Jones \& Forman (1992) \nocite{DJF92} look for X-ray emission,
detected by {\em Einstein}, associated with galaxies in the \IRAS
Bright Galaxy Sample.  In Fig.~\ref{fig:djf92}, we reproduce their
fig.~2, including the best-fit $L_X$ to $L_{\rm FIR}$ relationship,
which takes account of both detections and upper limits.  They find
that the X-ray luminosities of the Seyfert galaxies in the sample are
approximately ten times more X-ray luminous than normal or starburst
galaxies with the same FIR luminosity.

The X-ray/FIR relationship for R117 is also very similar to the spiral
galaxies studied by \nocite{Read97} Read, Ponman \& Strickland (1997),
where the diffuse fraction of the X-ray emission is of order
$\approxgt 50$ per cent, and is more typical of a normal starburst
galaxy than a merger \cite{Read98}.  Since both the \ISO $15\um$ and
radio emission are extended, we take this as evidence for the presence
of an extended starburst in R117\_A.  We therefore propose that at
least $50$ per cent of the X-ray emission is associated with the
starburst, but note that we cannot rule out some contribution from an
obscured AGN at a level of $\approxlt 50$ per cent.

\subsection{Sub-mm spectral index}

By measuring the apparent spectral index below the peak of the thermal
emission, an indication of the opacity of the dust can be obtained.
Since a pure black-body satisfies the Rayleigh-Jeans law for $h\nu \ll
kT$, the flux density obeys the relationship $S_\nu \propto \nu^2$.
For optically thin dust, the spectrum is much steeper, and $S_\nu
\propto \nu^{2+\beta}$, where usually $1<\beta<2$ (see 
\eg ~\citeNP{Cimatti97}).  Using the SCUBA $850\um$ flux and $450\um$
upper limit, we calculate that $\beta \le 0.86$ for R117, implying
that the dust here is almost optically thick.  However, the presence
of dust at a range of temperatures will have the effect of lowering
the measured value of $\beta$, which may be the case here.

\subsection{Radio/FIR properties}

Following the method of \citeANP{Carilli99} (1999), we can use the
radio to sub-mm flux ratio of R117 as a diagnostic of the emission
mechanism of the low frequency radiation in this galaxy.  In their
paper, \citeANP{Carilli99} state that for a starburst galaxy, both the
radio emission (1.4\GHz) and sub-mm emission ($850\um \equiv 350\GHz$)
are directly proportional to the total star-formation rate, and
therefore that their ratio, $S_{350}/S_{1.4}$, will be constant in the
source frame.  However, since the spectral shape changes slope between
these two regimes (\eg ~see Fig.~\ref{fig:sed}), then the {\em
observed} spectral index:

\[
\alpharsm = 0.42 \log_{10} \left( \frac{S_{350}}{S_{1.4}} \right),
\]

\ni will evolve with redshift due to the $k$-correction effect.  For a
source of unknown redshift, measurement of $\alpharsm$ can then be
used to place limits on the redshift.  Conversely, if the redshift is
known, then any dominant emission component from something other than
star-formation, \ie, the presence of an AGN, will serve to move the
galaxy away from the predicted position in the $\alpharsm$ 
\vs ~redshift plane.

R117 has flux densities of $S(1.4\GHz) \sim 1.4\mJy$ and $S(350\GHz)
\sim 4 \mJy$, which gives $\alpharsm \sim 0.19^{+0.049}_{-0.056}$.
Knowing that this galaxy lies at a redshift of $z=0.061$, we compare
the value of $\alpharsm$ with the predicted values, and find that R117
is totally consistent with a starburst galaxy at this redshift.  In
this scenario, the observed flux is due to synchrotron emission from
supernova remnants associated with the starburst \cite{Condon92}.  

We cannot rule out the presence of an AGN with this method, but can
infer that any AGN component will not be the dominant emission
mechanism at these wavelengths.

\subsection{Comparison with other galaxies}

\bfig
\centerline{
\psfig{figure=r117_comparison.ps,width=3.0in,angle=270}}
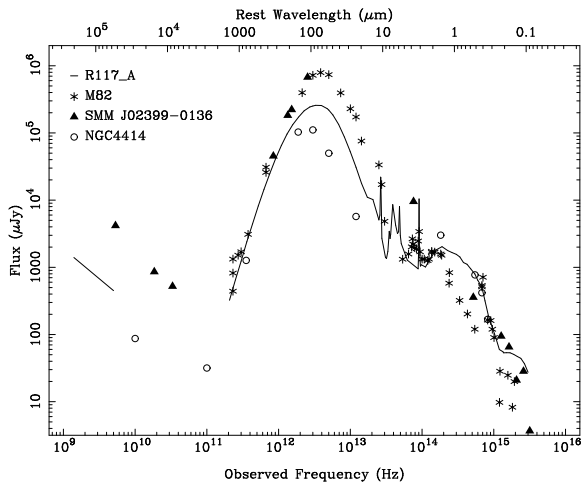
\caption{The spectral energy distribution of R117\_A compared with that
of various prototypical galaxies.  The model fit from Fig.~\ref{fig:sed}
is plotted as a solid line, and the comparison galaxies are plotted as
symbols as labelled.  M82 is a nearby strongly star-forming galaxy;
SMM\,J02399-0136 is a sub-mm selected, gravitationally lensed,
high-$z$, dusty AGN; and NGC\,4414 is a typical `normal' spiral
galaxy.  The data are plotted in the rest-frame of each galaxy, and
are normalized to the far infra-red flux of R117\_A, except in the
case of NGC\,4414, which is normalized to the optical flux of
R117\_A.}
\label{fig:comparison}
\efig

In Fig.~\ref{fig:comparison}, we compare the model fit to the spectral
energy distribution (SED) of R117 with the SEDs of a `normal' galaxy,
and those of infra-red luminous galaxies at both low and high
redshift.

When the SED of the typical `normal' spiral galaxy, NGC\,4414
\cite{Braine99}, is normalized to the optical flux of R117, we find
that the FIR flux of R117 is a factor of five times higher than that
of NGC\,4414, implying considerably more star-formation activity in
R117.  However, as seen in Fig.~\ref{fig:comparison}, R117 is not as
strongly star-forming as extreme examples of starburst galaxies such
as M82, implying an intermediate star-formation rate, between those of
M82 and NGC\,4414.  Arp\,220, the dusty AGN F10214+4724, the extremely
red object HR10, and M82, would all be underluminous in the optical
for the same FIR luminosity as R117.

Fig.~\ref{fig:comparison} also shows that R117 has a similar SED to
the sub-mm selected high redshift galaxy, SMM\,J02399-0136
\cite{Ivison98}, which is interesting as this galaxy is not only
strongly star-forming, but also contains an active nucleus.  However,
the SED of SMM\,J02399-0136 is not strongly defined in the mid
infra-red part of the spectrum.

\subsection{Alternative optical counterparts}
\label{sec:altopt}

Although in Section \ref{sec:select} we have shown on statistical
grounds that the probability of the galaxy R117\_A being a chance
identification is low, we nonetheless examine the optical, infra-red
and radio images of the \PSPC X-ray error circle for any indication of
a likely alternative identification.  Examination of
Fig.~\ref{fig:phot} shows that no objects of unusual colour lie in or
very close to the errorbox.  Neither are there any bright objects in
or near the errorbox.

The nearest bright object is a stellar object 20 arcsec from the X-ray
centroid to the west, R117\_B, which is visible on the $B$, $R$ and
$K$ images.  Recent spectroscopy has identified R117\_B to be an
M-star, with a magnitude of $R=18.7\pm0.2$.  M-stars often show
coronal X-ray emission and for the typical X-ray/optical ratios found
by \citeANP{Stocke91} (1991), such an M-star could be a noticeable
contributor to the total X-ray flux of R117.  Its location, far from
the X-ray centroid, rules out the M-star as being the major
contributor but is very consistent with it contributing enough flux to
pull the X-ray centroid a little away from R117\_A.

The two faint objects seen on the edge of the error circle in the
\bband image have $B\sim 25$ and $R>24$.  Even if identified as
unobscured QSOs, such faint objects would be unlikely to account for
more than a small fraction of the X-ray flux of R117 (see fig.~9,
\citeNP{McHardy98}).

\bfig
\centerline{
\psfig{figure=r117_obsqso.ps,width=3.0in,angle=270}}
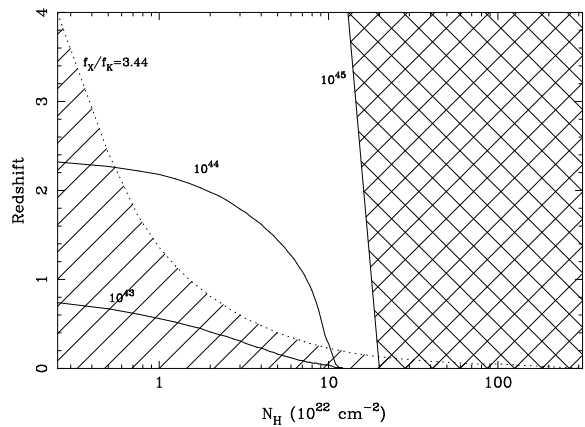
\caption{The region of $(N_H,z)$ parameter space for a putative
obscured QSO allowed by the observed soft X-ray flux and \kband limit
for R117.  The solid lines show the position of QSOs with unabsorbed
$0.5-2\keV$ luminosities as labelled (in $\fluxerg$), which for a
certain column density and redshift would give a $0.5-2\keV$ flux
equal to that observed for R117.}
\label{fig:obsqso}
\efig

The deep UKIRT/UFTI image (4050s) shows no evidence for any faint
objects closer to the X-ray centroid than R117\_A to a 5$\sigma$ limit
of $K\sim 20.8$.  This limit is fainter that the magnitudes of any
obscured QSOs found so far in \ROSAT surveys \cite{Lehmann2000}, and
so argues against a large contribution to the X-ray flux of R117 from
such a QSO.  However it is never possible to completely exclude an
absorbed QSO from any X-ray errorbox, including from \HRI errorboxes
already containing bright QSOs, and so we consider which regions of
parameter space are allowable for an obscured QSO.
Fig.~\ref{fig:obsqso} shows the region of $(N_H,z)$ parameter space
allowed by these observations.  The hatched region excludes objects
which would be too bright in $K$ for the observed \ROSAT flux, i.e.,
$f_X/f_K < 3.44$.  The cross-hatched region excludes objects with $N_H
\approxgt 2 \times 10^{23} \col$, which would not have been detected
by \ROSATc, even if at high redshift.  The solid lines show the
position in the $N_H$ {\em vs} $z$ plane of QSOs with unabsorbed
luminosities of $L_X (0.5-2\keV) = 10^{43}, 10^{44}, 10^{45} \ergps$
for which the received \ROSAT flux is as detected for R117.

We can therefore see that the only allowed luminosity range for an
absorbed QSO is between $\sim$few $\times 10^{43}$ and
$10^{45}\ergps$.  As discussed above, such a QSO cannot be ruled out
from almost any X-ray errorbox but, by Occam's razor, the bright
galaxy R117\_A is a much more likely identification.  If an obscured
QSO, with carefully tuned redshift and absorption, did contribute to
the X-ray flux from R117, we would have to explain why the X-ray flux
from R117\_A was less than that from other similar starburst galaxies.

\section{Conclusions}
\label{sec:concl}

Here we have presented a multiwavelength study of the properties of
the \ROSAT X-ray source R117, from the radio to the X-ray.  Having
ascertained that the $R=15.9$ galaxy, R117\_A, is the most likely
source of the X-ray emission, we have used a variety of techniques in
order to determine the dominant process behind this activity.  The SED
of the galaxy is well modelled by a spiral galaxy with a moderate
burst of star-formation.  Optical photometry and morphology are
consistent with R117\_A being a spiral galaxy, with a central bar
structure or nuclear star-forming region.  This galaxy counterpart to
the X-ray source is highly plausible on positional grounds, and in the
absence of any strong alternatives.  The ratios between the X-ray, FIR
and radio fluxes are consistent with this emission being due to a
starburst, as if an AGN were the dominant source of the power in this
object, the X-ray flux would be ten times larger.  Spectropolarimetric
observations of R117\_A place strong limits on the presence of any
broad components to the optical emission lines which would be expected
from AGN activity.  The starburst nature of R117 is also consistent
with both the detection of weak radio emission and the radio-to-submm
spectral index.  From the multiwavelength photometric data, combined
with the lack of any broad emission line component in polarized light,
we therefore conclude that starburst activity is the dominant source
of the {\em soft} X-ray flux detected by \ROSAT from the galaxy R117.

The broader implications of these observations concern the
identification and classification of X-ray sources found in deep X-ray
surveys.  In particular, galaxy identifications have been criticised
as being misidentifications in \ROSAT X-ray surveys and it has been
claimed that there is no starburst contribution to deep X-ray surveys
(e.g. \citeNP{Lehmann2000}).  Here we show that, in at least one case,
a starburst galaxy is by far the most likely identification of a faint
\ROSAT X-ray source.  We also note that \citeANP{Fiore2000} (2000)
find one definite starburst galaxy in a sample of a dozen \Chandra
sources.  Although we agree with the hypothesis that obscured AGN are
probably major contributors to the XRB, and that high luminosity NELGs
are probably mostly powered by an AGN, we conclude that low luminosity
X-ray emission from starburst galaxies cannot be entirely neglected as
contributing to the X-ray background, particularly at softer energies.

\section*{Acknowledgments}
 
We would like to thank the staff of the WHT, CFHT, UH 88-in, UKIRT,
JCMT and VLA telescopes and their observatories for their help in
obtaining the data on R117, and in particular, the UKIRT Service
Program for their prompt observation of R117.  We thank Harry Lehto
for assistance with the $6\cm$ VLA observations, Rob Ivison for the
SED datapoints of M82, etc., and Mat Page for useful discussions.  KFG
and IMH thank PPARC for support.  This paper was prepared using the
Southampton STARLINK node facilities.

\bsp

\label{lastpage}

\end{document}